\documentstyle[11pt,aaspp4,flushrt,epsfig]{article}

\begin{document}

\title{Differences in the Cooling Behavior of Strange Quark Matter Stars and
    Neutron Stars}
\author{Christoph Schaab\altaffilmark{1}, Bernd Hermann, 
        Fridolin Weber and Manfred K. Weigel} 
\affil{Institut f{\"u}r theoretische Physik,
  Ludwig-Maximilians Universit{\"a}t M{\"u}nchen, Theresienstr. 37,
  D-80333 M{\"u}nchen, Germany}
\altaffiltext{1}{E-mail address: schaab@gsm.sue.physik.uni-muenchen.de}

\begin{abstract}
The general statement that hypothetical strange (quark matter) stars
cool more rapidly than neutron stars is investigated in greater
detail. It is found that the direct Urca process could be forbidden
not only in neutron stars but also in strange stars. In this case,
strange stars are slowly cooling, and their surface temperatures are
more or less indistinguishable from those of slowly cooling neutron
stars.  Furthermore the case of enhanced cooling is reinvestigated.
It shows that strange stars cool significantly more rapidly than
neutron stars within the first $\sim 30$ years after birth. This
feature could become particularly interesting if continued observation
of SN 1987A would reveal the temperature of the possibly existing
pulsar at its center.
\end{abstract}


\section{Introduction}
The theoretical possibility that strange quark matter -- made up of
roughly equal numbers of up, down and strange quarks -- may be more
stable than atomic nuclei (specifically iron, which is the most stable
atomic nucleus) constitutes one of the most startling predictions of
modern physics (see
\cite{Bodmer71,Witten84,Terazawa}), which,
if true, would have implications of greatest importance for laboratory
physics, cosmology, the early universe, its evolution to the present
day, and massive astrophysical objects (cf. \cite{Aarhus91}).
Unfortunately it seems unlikely that lattice QCD calculations will be
accurate enough in the foreseeable future to give a definitive
prediction on the absolute stability of strange matter, so that one is
presently left with experiments and astrophysical studies
(cf. \cite{Glendenning92,Glendenning94b}) to either confirm or reject the
absolute stability of strange matter. This letter, dealing with the
second item, compares the cooling behavior of neutron stars with the
one of their hypothetical strange counterpars -- strange stars
(cf. \cite{Witten84,Haensel86,Alcock86,Glendenning90}).  The
theoretical predictions are compared with the body of observed data
taken by ROSAT and ASCA. There have been investigations on this topic
prior to this one (e.g., see
\cite{Alcock88,Pizzochero91,Page91a,Schaab95a}). These, however, did not
incorporate the so-called standard cooling scenario that turns out to
be possible not only in neutron star matter but in strange quark
matter too, altering some of the conclusions made in the earlier
investigations significantly.


\section{Description of strange matter} 
We use the MIT bag model including $O(\alpha_{\rm{c}})$-corrections
(cf. \cite{Chados74,Farhi84}) to model the properties of absolutely
stable strange matter. Its equation of state and quark-lepton
composition, which is governed by the conditions of chemical
equilibrium and electric charge neutrality, is derived for that range
of model parameters, that is, bag constant $B^{1/4}$, the strange
quark mass $m_{\rm{s}}$, and strong coupling constant
$\alpha_{\rm{c}}$, for which strange matter is absolutely stable
(energy per baryon $E/A$ less than the one in iron, $E/A=930$~MeV). In
the limiting case of vanishing quark mass, the electrons are not
necesaary to maintain charge neutrality. In the realistic case of
finite strange quark masse $m_{\rm s}$, the electrons can nevertheless
vanish above some density which depends on $\alpha_{\rm c}$. It was
pointed out by Duncan et al. (1983) (see also \cite{Alcock86,Pethick92a})
that the neutrino emissivity of strange matter depends strongly on its
electron fraction, $Y_{\rm e}$. For that reason we introduce two
different, complementary parameter sets denoted SM-1 and SM-2 (see
Tab.  \ref{tab:parameter}), which correspond to strange matter that
contains a relatively high electron fraction (SM-1), and $Y_{\rm e}=0$
(SM-2) for the density range being of interest here.

\placetable{tab:parameter}


\section{Neutrino emissivity} 
The quark {\em direct} Urca processes
\begin{equation}
  \rm{d} \rightarrow \rm{u} + \rm{e}^- + \bar\nu_{\rm{e}} \label{eq:dtou}
\end{equation}
and
\begin{equation}
  \rm{s} \rightarrow \rm{u} + \rm{e}^- + \bar\nu_{\rm{e}}, \label{eq:stou}
\end{equation}
as well as their inverse ones are only possible if the fermi
momenta of quarks and electrons ($p_{\rm{F}}^i$, $i=$u,d,s;e$^-$) fulfill the
so-called triangle inequality (e.g.,
$p_{\rm{F}}^{\rm{d}}<p_{\rm{F}}^{\rm{u}}+p_{\rm{F}}^{\rm{e}}$ for process
(\ref{eq:dtou})). This relation is the analogue to the triangle inequality
established for nucleons and electrons in the nuclear matter case (direct Urca
process, \cite{Boguta81a,Lattimer91}).

If the electron fermi momentum is too small (i.e., $Y_{\rm e}$ is too little),
then the triangle inequality for the above processes (\ref{eq:dtou}) and
(\ref{eq:stou}) cannot be fulfilled, and  a bystander quark is needed to
ensure energy and momentum conservation in the scattering process. The latter
process is known as the quark {\em modified} Urca process, whose emissivity is
considerably smaller than the emissivity of the direct Urca process.  If the
electron fraction vanishes entirely, as is the case for SM-2, both the
quark direct and the quark modified Urca processes become unimportant. The
neutrino emission is then dominated by bremsstrahlung processes only,
\begin{equation}
  Q_1 + Q_2 \longrightarrow Q_1 + Q_2 + \nu + \bar\nu,
\end{equation} where $Q_1$, $Q_2$ denote any pair of quark flavors.  For the
emissivities associated with the quark direct Urca, quark modified
Urca, and quark bremsstrahlung processes, we refer to 
Refs. Price (1980), Iwamoto (1982) and Duncan et al. (1983).

It has been suggested (see \cite{Bailin79a,Bailin84}) that the quarks
eventually may form Cooper pairs. This would suppress, as in the
nuclear matter case, the neutrino emissivities by an exponential
factor of $\exp(-\Delta/k_{\rm{B}}T)$, where $\Delta$ is the gap
energy, $k_{\rm{B}}$ Boltzmann's constant, and $T$ the
temperature. Unfortunately, there exists up to now neither a precise
experimentally nor theoretically determined value of the gap energy.
So to provide a feeling for the influence of a possibly superfluid
behavior of the quarks in strange matter too, we choose
$\Delta=0.4$~MeV as estimated in the work of Bailin \& Love (1979). (Such a
$\Delta$ value is not too different from the nuclear-matter case,
where the proton $^1{\rm S}_0$ gap, for instance, amounts $\sim 0.2 
{\rm -} 1.0$~MeV (cf. \cite{Wambach91a,Elgaroy96c}), depending on the
nucleon-nucleon interaction and the microscopic model.) The outcome of
our superfluid strange matter calculations will be labeled
SM-1$^{\rm{sf}}$ and SM-2$^{\rm{sf}}$.

\section{Observed data}
Among the X-ray observations of the 14 sources which
were identified as pulsars, the ROSAT and ASCA observations of PSRs 0833-45
(Vela), 0656+14, 0630+18 (Geminga) and 1055-52 (see Tab.
\ref{tab:observations}) achieved a sufficiently high photon flux such that the
effective surface temperatures of these pulsars could be extracted by
two- or three-component spectral fits (cf. \cite{Oegelman95a}).  The
obtained effective surface temperatures, shown in Figs.  \ref{fig:nsf}
and \ref{fig:sf}, depend crucially on whether a hydrogen atmosphere is
used or not.  Since the photon flux measured solely in the X-ray
energy band does not allow one to determine what kind of atmosphere
one should use, we consider both the blackbody model and the
hydrogen-atmosphere model, drawn in in Figs.  \ref{fig:nsf} and
\ref{fig:sf} as solid and dashed error bars, respectively.  The kind
of atmosphere possessed by a specific pulsar could be determined by
considering multiwavelength observations (see \cite{Pavlov96a}). All error
bars represent the $3\sigma$ error range due to the small photon
fluxes. The pulsars' ages are determined by their spin-down times
assuming a canonical value of 3 for the braking index. In reality the
braking index may be quite different from 3. Its variation between 2
and 4, for instance, would change the age of Geminga as indicated by
the horizontal error bar shown at the bottom of Figs. \ref{fig:nsf}
and \ref{fig:sf}.

\placetable{tab:observations}


\section{Results and Discussion} 
The thermal evolution of strange stars and neutron stars was simulated using
the evolutionary numerical code described in Schaab et al. (1996b) (see also
\cite{Tsuruta66,Richardson82,VanRiper91,Page95,Schaab95b}). The neutron star models are based on a broad collection of
EOSs which comprises relativistic, fieldtheoretical equations of state as well
as non-relativistic, Schroedinger-based ones (see \cite{Schaab95a} for
details).  As a specific feature of the relativistic models, they account for
all baryon states that become populated in dense neutron star matter up to the
highest densities reached in the cores of the heaviest neutron stars
constructed from this collection of equations of state.  Neutron stars are
known to loose energy either via standard cooling or enhanced cooling.  Both
may be delayed by superfluidity.  Consequently all four options have
been taken into account here. These are labeled in Figs.  \ref{fig:nsf} and
\ref{fig:sf} as NS-1 (enhanced cooling) and NS-2 (standard cooling) for normal
neutron star matter, and NS-1$^{\rm{sf}}$ and NS-2$^{\rm{sf}}$ (delayed
cooling) for superfluid neutron star matter.  The parameters of
NS-1$^{\rm{sf}}$ and NS-2$^{\rm{sf}}$ are listed in Tab.~4 of Ref.
Schaab et al. (1996b).  In analogy to this, the corresponding strange-star cooling
curves are SM-1 (enhanced cooling) and SM-2 (standard cooling) for normal
strange quark matter, and SM-1$^{\rm{sf}}$ and SM-2$^{\rm{sf}}$ (delayed
cooling) for superfluid quark matter.

\placefigure{fig:nsf}
\placefigure{fig:sf}

All calculations are performed for a star mass of $M=1.4M_\odot$,
about which the observed pulsar masses tend to scatter. The band-like
structure of the cooling curves is supposed to reflect the
uncertainties inherent in the equation of state of neutron-star and
strange-star matter.  These have their origin, in the case of neutron
stars (bands filled with dots), in the different many-body techniques
used to solve the nuclear many-body problem, and the star's
baryon-lepton composition.  In the latter case, strange-star matter,
the solid bands refer to different bag values, $B^{1/4}$, which vary
from 137 to 148 MeV for SM-1, and from 133 to 146 MeV for SM-2. All
values correspond to absolutely stable strange matter.  One might
suspect that the large gap between the cooling tracks of the SM-1 and
SM-2 models in Fig. \ref{fig:nsf} can be bridged steadily by varying
the strong coupling constant $\alpha_{\rm c}$ in the range
0.1--0.15. However it turns out that the gap can be filled only for 
$\alpha_{\rm c}$-values within an extremely small range. This is caused
by the sensitive functional relationship between $\alpha_{\rm c}$ and
the neutrino luminosity $L_\nu$, which is rather steep around that
$\alpha_{\rm c}$-value for which the electrons vanish from the quark
core of the star. All other values of $\alpha_{\rm c}$ give cooling
tracks which are close to the upper or lower bands, respectively. This
behavior might be compared with the case of neutron stars, where the
neutrino luminosity depends sensitively on the star's mass.

One sees from Figs. \ref{fig:nsf} and \ref{fig:sf} that, except for the first
$\sim 30$ years of the lifetime of a newly born pulsar, both neutron stars and
strange stars may show more or less the {\em same} cooling behavior, provided
both types of stars are made up of either normal matter or superfluid matter.
(We will come back to this issue below.) This is made possible by the fact that
both standard cooling (NS-2) as well as enhanced cooling (NS-1) in neutron
stars has its counterpart in strange stars too (SM-2 and SM-1, respectively).
The point of time at which the surface temperature drop of a strange star
occurs depends on the thickness of the nuclear crust that may envelope the
strange matter core (cf. \cite{Schaab95a}). In the present calculation, strange stars
possess the densest possible nuclear crust, which is about 0.2 km thick.
Thinner crusts would lead to temperature drops at even earlier times, and thus
an earlier onset of the photon cooling era.  Figures \ref{fig:nsf} and
\ref{fig:sf} indicate that the cooling data of observed pulsars do not allow to
decide about the true nature of the underlying collapsed star, that
is, as to whether it is a strange star or a conventional neutron star.
This could abruptly change with the observation of a very young pulsar
shortly after its formation in a supernova explosion.  In this case a
prompt drop of the pulsar's temperature, say within the first 30 years
after its formation, could offer a good signature of a strange star
(see \cite{Alcock88,Pizzochero91}).  This feature, provided it
withstands a rigorous future analysis of the microscopic properties of
quark matter, could become particularly interesting if continued
observation of SN 1987A would reveal the temperature of the possibly
existing pulsar at its center.

Finally, we add some comments about the possibility that only the
neutron star is made up of superfluid matter but not the strange star.
In this case one has to compare the models SM-1 and SM-2 (see
Fig. \ref{fig:nsf}) with models NS-1$^{\rm sf}$ and NS-2$^{\rm sf}$
(see Fig. \ref{fig:sf}) yielding to an overall different cooling
history of neutron stars and enhanced-cooling strange stars
(SM-1). Therefore, the standard argument pointed out quite frequently
in the literature that strange stars cool much more rapidly than
neutron stars applies only to this special case.
  
\clearpage

\begin{table}
  \caption{Bag constant, $B^{1/4}$, strange quark mass, $m_{\rm{s}}$, and
    the QCD coupling constant, $\alpha_{\rm{c}}$, for the two sets of
    parameters denoted SM-1 and SM-2. The energy per baryon, $E/A$, for two and
    three flavor quark matter is given too. \label{tab:parameter}}
\centering \begin{tabular}{lcc}
  Quantity  & SM-1 & SM-2 \\
  \tableline
  $B^{1/4}$ [MeV]    & 140 & 140 \\
  $m_{\rm{s}}$ [MeV] & 150 & 150 \\
  $\alpha_{\rm{c}}$  & 0.10 & 0.15 \\
  $E/A$ in 2-flavor quark matter [MeV]& 959 & 987 \\
  $E/A$ in 3-flavor quark matter [MeV]& 878 & 892 \\
\end{tabular}
\end{table}

\begin{table}
  \caption[]{Surface temperatures, $T_{\rm s}^\infty$, and spin-down
  ages, $\tau$, of observed pulsars.} \label{tab:observations}
\centering \begin{tabular}{ccccc}
Pulsar & $\tau$ [yr] & Model atmosphere & $T_{\rm s}^\infty$ [K] 
& Reference \\
\tableline
0833-45 & $1.1\times 10^4$ & blackbody             
& $1.3\times 10^6$               & \cite{Oegelman93a} \\
(Vela)  &                  & magnetic H-atmosphere 
& $7.0^{+1.6}_{-1.3}\times 10^5$ & \cite{Page96a} \\
\tableline
0656+14 & $1.1\times 10^5$ & blackbody
& $7.8^{+1.5}_{-4.2}\times 10^5$ & \cite{Greiveldinger96a} \\
        &                  & magnetic H-atmosphere 
& $5.3^{+1.2}_{-0.9}\times 10^5$ & \cite{Anderson93} \\
\tableline
0630+18 & $3.2\times 10^5$ & blackbody
& $5.2\pm 3.0\times 10^5$        & \cite{Halpern93a} \\
(Geminga)&                 & H-atmosphere 
& $1.7\pm 1.0\times 10^5$        & \cite{Meyer94} \\
\tableline
1055-52 & $5.4\times 10^5$ & blackbody
& $7.9^{+1.8}_{-3.0}\times 10^5$ & \cite{Greiveldinger96a} \\
\end{tabular}
\end{table}


\clearpage
\figcaption[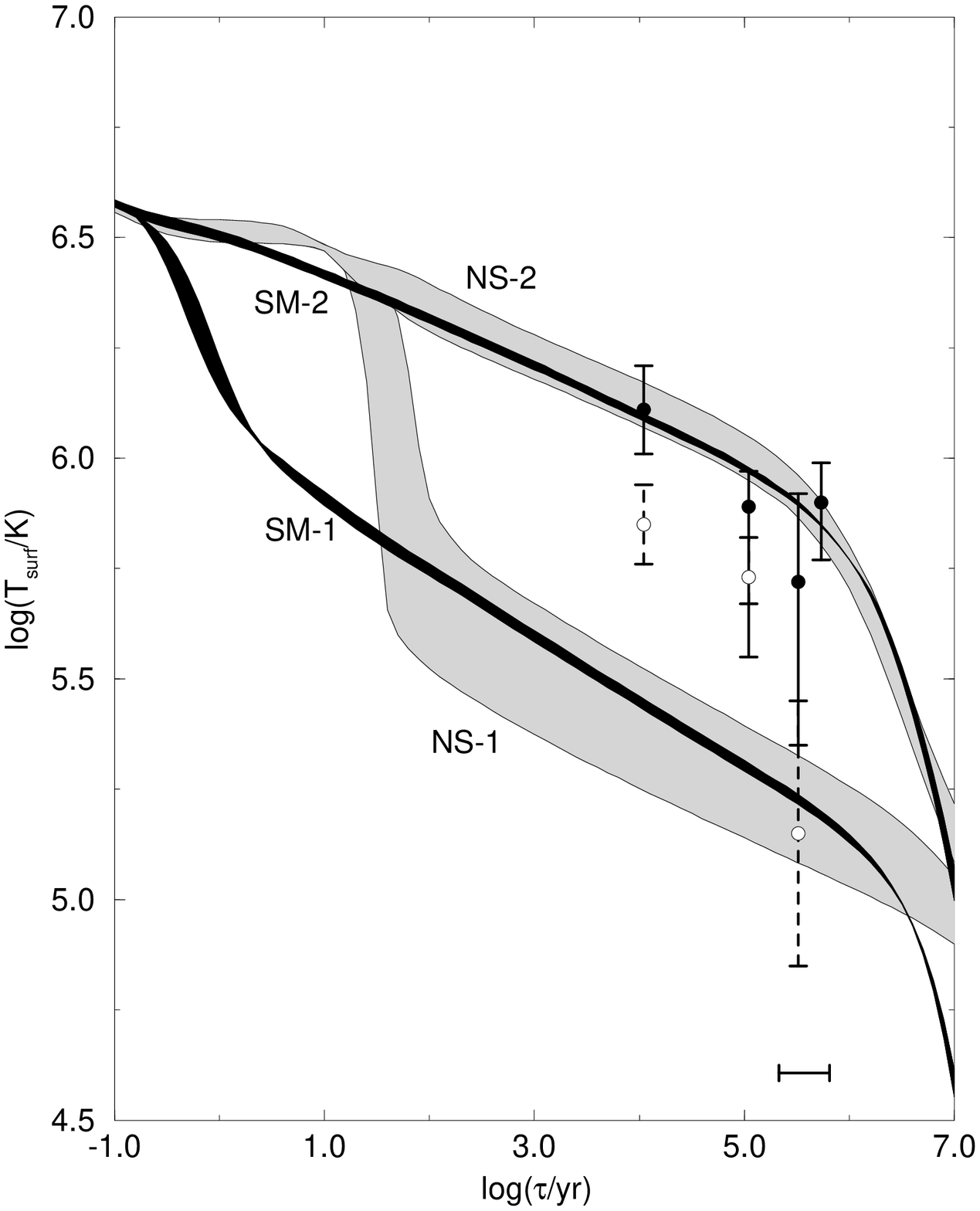]{
  Cooling of non-superfluid strange star models SM-1 (lower solid
  band) and SM-2 (upper solid band), and neutron star models NS-1
  (lower dotted band) and NS-2 (upper dotted band). The surface
  temperatures obtained with a blackbody- (magnetic hydrogen-)
  atmosphere are marked with solid (dashed) error bars (see
  Tab. \ref{tab:observations}). The uncertainty in the pulsar's age is
  displayed by the error bar at the bottom.  \label{fig:nsf}}

\figcaption[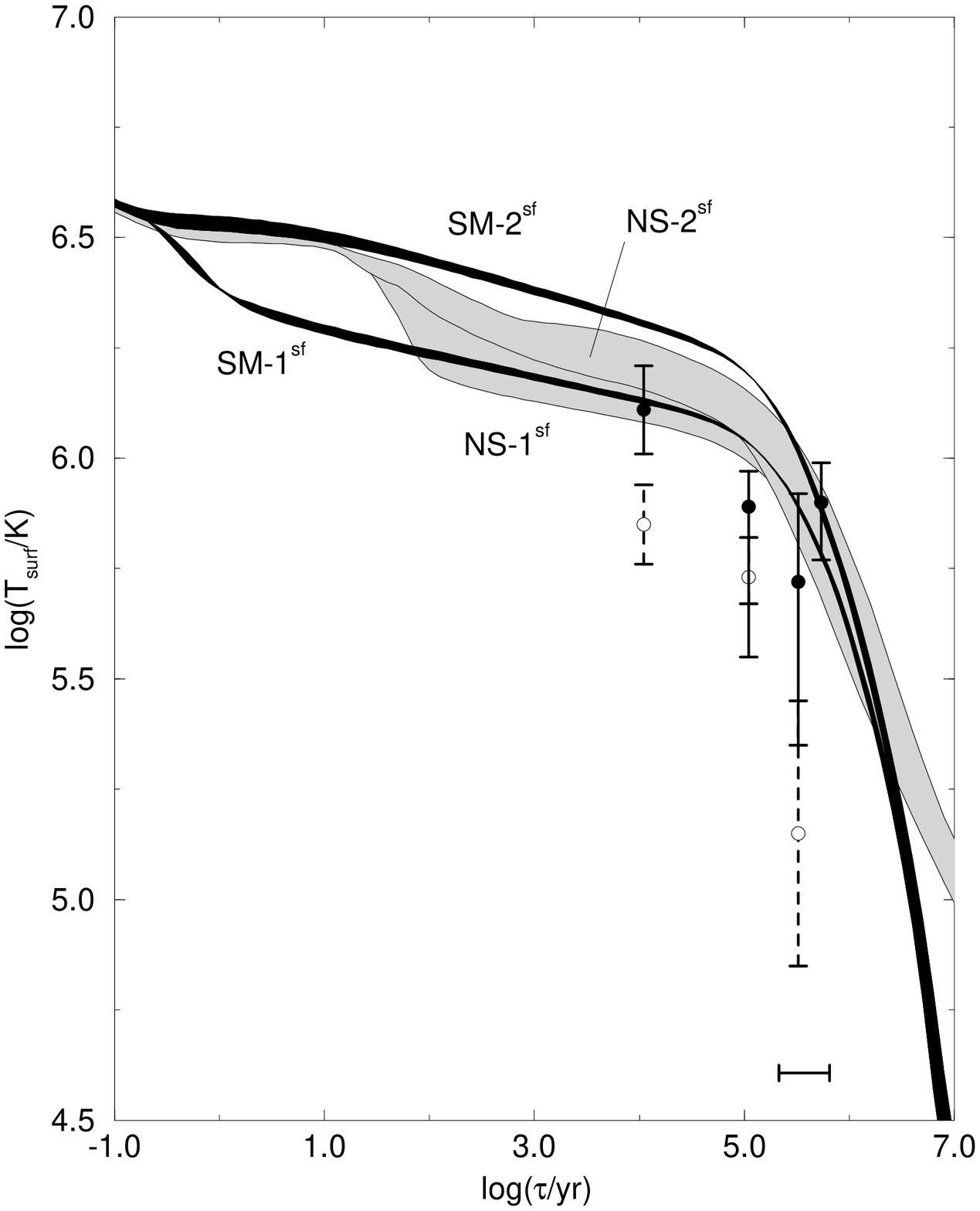]{
  Cooling of superfluid strange star SM-1$^{\rm sf}$ (lower solid 
  band) and SM-2$^{\rm sf}$ (upper solid band), and neutron star
  models NS-1$^{\rm sf}$ (lower dotted band) and NS-2$^{\rm sf}$ (upper
  dotted band). \label{fig:sf}}

\clearpage\vspace*{\fill}
\psfig{figure=p0696_1.ps,width=0.8\linewidth}
\par\vspace{2cm}\centering\large\bf Fig. 1
\clearpage\vspace*{\fill}
\psfig{figure=p0696_2.ps,width=0.8\linewidth}
\par\vspace{2cm}\centering\large\bf Fig. 2

\end{document}